\documentclass[a4paper,12pt]{article}

\usepackage{ifpdf}
\newif\ifpdf
\ifx\pdfoutput\undefined
  \pdffalse
\else
  \pdfoutput=1
  \pdftrue
\fi

\RequirePackage{xspace} %
\RequirePackage{subfigure} %
\RequirePackage[centertags]{amsmath} %
\RequirePackage{amssymb}
\RequirePackage{wrapfig} %
\RequirePackage{calc} %
\RequirePackage{ifthen}
\RequirePackage{tabularx} %
\RequirePackage{flafter} %
\RequirePackage{fancyhdr} %

\ifpdf
  \RequirePackage[pdftex]{color}%
  \RequirePackage{colortbl}%
  \RequirePackage{array}%
  \RequirePackage[pdftex]{graphicx}
  \RequirePackage[ pdftex, plainpages = false, pdfpagelabels,
                 pdfpagelayout = useoutlines,
                 bookmarks,
                 breaklinks = true,
                 linktocpage,
                 pagebackref,                      
                 colorlinks = true,
                 linkcolor = blue,
                 urlcolor  = blue,
                 citecolor = blue,
                 anchorcolor = blue,
                 hyperindex = true,
                 hyperfigures
                 ]{hyperref}
\else
  \RequirePackage{color}
  \RequirePackage{colortbl}
   \RequirePackage{array}
  \RequirePackage[dvips]{graphicx}
  \RequirePackage{hyperref}
  \usepackage{rotating}
\fi

\usepackage{makeidx} 
\usepackage{setspace} 
\usepackage{rotating} 
\usepackage{ecltree}
\usepackage{epic}
\usepackage{supertabular}  
\usepackage{color}
\usepackage{exscale}
\usepackage{fontenc}
\usepackage{ifthen}
\usepackage{latexsym}
\usepackage{makeidx}
\usepackage{syntonly}
\usepackage{inputenc}
\usepackage{graphicx}
\usepackage{setspace}
\usepackage{caption2}
\usepackage[english]{babel}
\usepackage[square, comma,numbers,sort&compress]{natbib}
\usepackage{hypernat}
\usepackage{boxedminipage}
\usepackage{framed}
\usepackage{longtable}
\usepackage{setspace}
\usepackage{caption2}
\usepackage{mflogo}
\usepackage{bbding}
\usepackage[all]{hypcap}

\usepackage{microtype}
\DisableLigatures{}

\newcommand{\note}[1]{\marginpar[left]{\singlespace \tiny #1}}

\renewcommand{\sectionmark}[1]%
      {\markright{\thesection\ #1}} 

\renewcommand{\note}[1]{}

\newcommand{\SA}     {\theta}                                                   
\newcommand{\etal}    {{\it et al}}

\doublespace 

\begin{document}


\title{\vspace* {2.0cm} New program with new approach for spectral data analysis \vspace{4.0cm}}

\author{Taha Sochi\footnote{University College London, Department of Physics \& Astronomy, Gower Street, London, WC1E 6BT.
Email: t.sochi@ucl.ac.uk.} \vspace*{5.0cm}}


\maketitle

\thispagestyle{empty} %


\pagenumbering{roman}

%
%

\pagestyle{headings} %
\addtolength{\headheight}{+1.6pt}
\lhead[{Chapter \thechapter \thepage}]%
{{\bfseries\rightmark}}
\rhead[{\bfseries\leftmark}]%
{{\bfseries\thepage}} 
\headsep = 1.0cm               %

\pagenumbering{arabic}

\newpage
\section{Abstract} \label{Abstract}

This article presents a high-throughput computer program, called EasyDD, for batch processing,
analyzing and visualizing of spectral data; particularly those related to the new generation of
synchrotron detectors and X-ray powder diffraction applications. This computing tool is designed
for the treatment of large volumes of data in reasonable time with affordable computational
resources. A case study in which this program was used to process and analyze powder diffraction
data obtained from the ESRF synchrotron on an alumina-based nickel nanoparticle catalysis system is
also presented for demonstration. The development of this computing tool, with the associated
protocols, is inspired by a novel approach in spectral data analysis.

Keywords: spectral analysis, synchrotron, X-ray, powder diffraction, high-throughput software,
tomographic imaging, multi-TEDDI detector, alumina-based catalyst.

\newpage
\section{Introduction}

The last few decades have witnessed a revolution in the detectors and data acquisition
technologies. This, associated with the computing and communication revolution, has increased the
demand for data processing power. Modern detectors coupled with the high intensity radiation
sources have led to the situation where data sets can be collected in ever shorter time scales and
in ever larger numbers. Such large volumes of data sets pose a data processing problem which
augments with the current and future instrument development.

EasyDD is based on a new approach for large-scale processing, visualization and analysis of massive
volumes of spectral data. Such a utility greatly assists studies on various physical systems and
enables far larger detailed data sets to be rapidly interrogated and analyzed. EasyDD methodology
is based on automation, batch processing, and encapsulation of various modules in a single entity
where the user can sequentially follow computational protocols to apply multiple types of
procedures on massive amounts of correlated data sets with general common features.

EasyDD can be described as a high throughput software to manage, process, analyze and visualize
spectral data in general and synchrotron data in particular. It is a powerful tool for processing
large quantities of data in a variety of formats with ease and comfort using limited time and
computing resources. The main features of EasyDD which were observed in its development are

\begin{itemize}

\item
User friendliness to minimize the time and effort required to learn and use. A graphic user
interface is therefore adopted in favor of command line interface although the latter is more
common in scientific computing and much easier to develop.

\item
Capability of handling several common data formats including a generic XY format so that it is
possible to use in processing data produced by various sensors and detectors.

\item
Optimization for the commonly available computational resources; most importantly CPU time and
memory. The program therefore tries to set the limits of its data processing capabilities to the
limits of the available computational resources.

\item
Batch and multi-batch processing functionalities which most EasyDD procedures are based upon.

\end{itemize}

These features allow processing huge amounts of data ($\sim$ terabytes) in reasonable time ($\sim$
hours or days).

EasyDD uses a hybrid approach of procedural and object oriented programming methodologies. It
combines Graphic User Interface (GUI) technology with standard scientific computing techniques. Its
resources include the standard C++ library, a GUI library, with numerous algorithms, functions and
techniques. Several input data formats are supported, which include but not limited to: generic XY,
MCA of Diamond synchrotron \cite{Diamond}, MCA of ESRF \cite{ESRF}, ERD of Manchester University
detector, and HEXITEC detector (e.g. \cite{HEXITEC}).

One of the main functionalities of EasyDD is to read and map spectral data on a graphic interface.
This can be done simply by depositing the data files of a particular format in a directory and
invoking the relevant read function. On reading the files, the data are stored in memory and mapped
on a 2D color-coded tab. Multiple tabs from different data sources can be created at the same time.
The tabs can also be removed collectively or individually in any order. In the following section we
outline the main components of EasyDD, with a brief account of their main functionalities, followed
by a general description of its modules.

\subsection{Components and Modules of EasyDD}

The principal component of EasyDD is the main window which is a standard GUI widget with menus,
toolbars, a status bar, context menus and so on. The basic functionality of this widget is to serve
as a platform for accessing and managing other components with their specific functionalities.

Another component is the tab widget which can accommodate a number of 2D color-coded scalable tabs
for tomographic mapping with graphic and text tooltips to show all essential data properties. The
tabs can be used to launch a plotter for dynamic display of individual patterns of the mapped data.
The tabs can also be used for imaging, 3D visualization, manipulation and format conversion of
these data.

A third component is a numeric plotter to obtain a graph of the spectral pattern for any cell in
the tabs. It is also used to create basis functions and forms for curve-fitting. The 2D plotter
capabilities include creating and drawing fitting basis functions which include polynomials of
order $\leq$ 6 that pass through a number of selected points, Gauss, Lorentz and pseudo-Voigt. The
fitting basis functions can also be modified and removed from within the plotter. The plotter can
be used to perform non-linear least-squares curve-fitting by Levenberg-Marquardt algorithm on
individual data in the tabs. Moreover, the plotter image can be saved in a number of different
formats.

A fourth component is a spreadsheet form, mainly used for batch curve-fitting. The idea is to
prepare a form in the plotter and save it to the disc. It is then imported for batch fitting a
number of cells or tabs in the tab widget or to use it in a multi-batch curve-fitting operation.
The form has a number of columns that contain data required for curve-fitting such as data range,
initial fitting parameters, upper and lower limits, boolean flags and values for applying
restrictions on the refined parameters when they exceed acceptable limits. Columns that contain
counters and boolean flags for controlling the number of iteration cycles and the parameters to be
refined in the least-squares fitting routine can also be added by the user.

A fifth component is a 3D plotter for creating a 3D graph of the data in the tabs where the total
intensity is displayed against the tomographic dimensions. The 3D plotter is very useful for close
inspection of data as the graph can be rotated in all orientations and zoomed in and out.

EasyDD contains four main modules; which are

\begin{itemize}

\item
Curve-fitting by least-squares minimization using Levenberg-Marquardt algorithm which is an
iterative nonlinear least-squares optimization numerical technique. Thanks to its efficiency and
good convergence, the Levenberg-Marquardt algorithm is widely used by scientists and engineers in
all disciplines, and hence it became a standard for nonlinear least-squares minimization problems.
In EasyDD, curve-fitting can be performed on a single pattern, or as a single batch process over
multiple patterns, or as a multi-batch process over multiple forms and data sets. Curve-fitting can
be done on a single or multiple peaks using a number of basis functions with and without polynomial
background modeling. The number of curve-fitting cycles can be fixed or vary according to the
convergence criteria. The parameters to be fitted can also be selected with possible application of
restrictions. The range of data to fit can be selected graphically or by using a prepared form.
Some relevant statistical indicators for the fitting process are computed in the curve-fitting
routine.

\item
Sinogram treatment and back projection which is a tomographic technique for reconstructing a 2D
image in real space from a number of 1D projections and their angles. It relies on the application
of the inverse Radon transform and the Fourier Slice Theorem. The raw data for back projection are
sinograms consisting of a set of rotation versus translation measurements. Back projection can run
in single and multi-batch modes and can be applied on total or partial intensity as well as
individual channels with possible application of Fourier transform and filtering.

\item
EDF processing to extract the information from EDF binary files obtained from CCD detectors
\cite{SochiEDFPaper2011}. Two forms of extraction are available: conversion to normal 2D images in
one of three formats (png, jpg and bmp), and squeezing to 1D patterns in $xy$ text format with the
possibility of making tilt and missing-ring corrections.

\item
Graphic presentation which includes mapping, visualization and imaging to produce graphs in 2D and
3D spaces. These include creating tomographic and surface images in single, batch and multi-batch
modes, as well as $xy$ plots for spectral patterns. Some of these graphic techniques use a direct
display on the computer monitor, while others save the results to the computer storage in the form
of image files.

\end{itemize}

\section{Case Study} \label{CaseStudy}

EasyDD has been used in a number of key studies such as \cite{LazzariJSB2009, EspinosaOJBJe2009,
SochiThesisBBK2010, JacquesMBSOe2011, LazzariEJSMe2012, ObrienJMBWB2012, GibsonZJBCe2013}. It has
also been used by the High Energy X-Ray Imaging Technology (HEXITEC) project for the development of
multi-pixel 2D X-ray detectors \cite{HEXITEC, CernikHMPAe2010, WilsonBCHJe2011}. However, in this
section we present a brief account of a nickel chloride catalysis system study as a show case for
the use of EasyDD in processing and analyzing powder diffraction data. The nickel chloride data,
which were collected by S. Jacques and coworkers (refer to \cite{SochiThesisBBK2010} for details),
are part of a larger data collection which consists of about 254 thousand EDF image files in 179
data sets (sinograms) with a total size of about 2.45 terabytes. The measurements were carried out
at the ESRF \cite{ESRF} beamline ID15B which is dedicated to applications using very high energy
X-ray radiation up to several hundreds of keV. ID15B houses the angle dispersive diffraction setup
using a large area detector and high resolution Compton spectrometer. The wavelength of the
monochromatic beam used in these measurements is $\lambda=0.14272$ \AA. The nickel chloride
collection contains 23 data sets representing sequential time frames of a single lateral slice in a
cylindrical object.

A computer aided tomography (CAT) technique in angle dispersive diffraction (ADD) mode was employed
in these measurements to monitor the chemical and crystallographic developments in a slice through
an impregnated extrudate sample undergoing heat treatment to obtain 2D information at various
points in time. A charge-coupled device (CCD) was used to record the diffraction patterns in 2D
space. The CAT type ADD method has been suggested previously in the literature and has recently
been demonstrated by Bleuet \etal\ \cite{BleuetWDSHW2008}. The temporal aspect of the study makes
it entirely novel and challenging, requiring the data acquisition to be sufficiently fast to make
the process observable.

A major advantage of using CAT type ADD technique is the rapid rate of data collection.
Proportional area detectors, such as CCD devices, when used for recording angle dispersive
diffraction patterns can support much higher count rates than energy dispersive solid state
detectors. Since these devices offer fast reading, they can provide a more thorough insight into
the temporal dimension of the dynamic processes under investigation. The area aspect of such
detectors allows the recording of entire powder diffraction rings of the whole diffraction pattern
simultaneously with good signal-to-noise ratio. The recorded intensity would be severely reduced if
only a strip detector was employed. Such area detectors, when used in conjunction with very bright
sources, allow for very fast data acquisition even from materials that give fairly poor scattering.

\begin{figure} [!h]
\centering
\includegraphics
[scale=0.9] {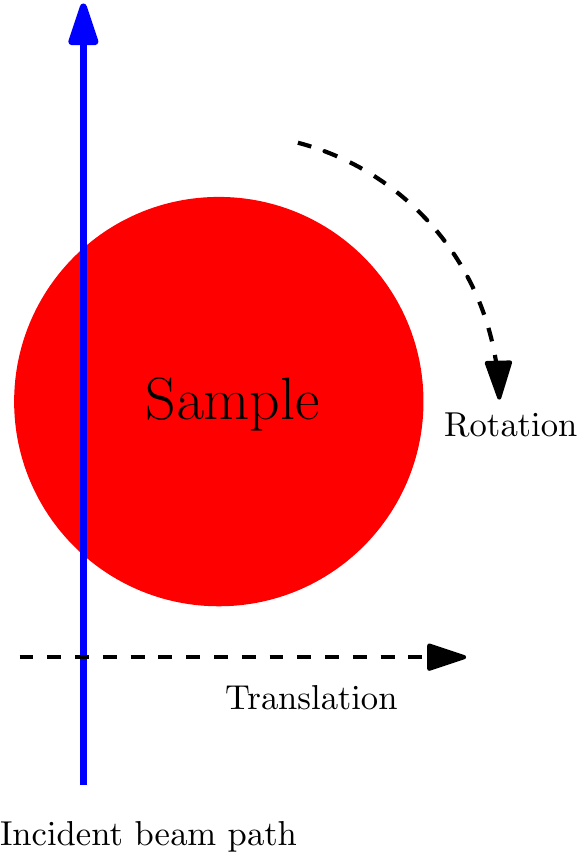}
\caption[Schematic demonstration of the scanning employed in the spatial/temporal CAT type ADD experiment.]%
{Schematic demonstration of the scanning employed in the spatial/temporal CAT type ADD experiment.}
\label{CAT}
\end{figure}

In the CAT technique, a pencil beam of monochromatic synchrotron X-ray is applied on a sample
mounted on translational-rotational stage and a time frame of the slice is collected for each
translation-rotation cycle, as depicted schematically in Figure \ref{CAT}. In each frame, the
sample is translated $m$ times across the beam and a complete diffraction pattern is collected for
each translation position. These $m$ translations are then repeated at $n$ angles between 0 and
$\pi$ in steps of $\pi/(n-1)$, and hence $m\times n$ diffraction patterns are collected for each
time frame. The complete data of a frame represent a sinogram that can be reconstructed, using a
back projection computational algorithm, to obtain a tomographic image of the slice in real space.
A series of frames then give a complete picture of the dynamic transformation of the phases
involved during the whole experiment. As area detectors are employed in these experiments, the 2D
diffraction images should be transformed to 1D patterns by integrating the diffraction rings.
Curve-fitting can then be used to identify the phases in each stage as the peaks in these patterns
provide distinctive signatures of each phase.

The sample used in this study is a cylindrical extrudate of $\gamma$-alumina (Al$_{2}$O$_{3}$) as a
base impregnated with nickel nanoparticles as an active metallic catalyst. The impregnation was
performed using an aqueous solution of nickel chloride ethylenediamine tetrahydrate,
NiCl$_2$(en)(H$_2$O)$_4$, as a precursor. The sample and preparation process are similar to those
described by Beale \etal\ \cite{BealeJBBW2007}. During the experiment, the sample was undergoing a
heat treatment which consists of ramp increasing temperature from 25$^\circ$C at the first frame to
500$^\circ$C at the 20th frame (i.e. 25$^\circ$C increase per frame) followed by steady state
temperature of 500$^\circ$C at the last three frames.

EasyDD was used in multi-batch mode to convert the binary EDF images to 1D spectral patterns in
ASCII numeric format. It was also used to visualize, align, and back project the sinograms; and
curve-fit the peaks of the back-projected patterns. A Gaussian profile was used to model the peak
shape while a linear polynomial was used to model the background. Most peaks were fitted as
singlets while the remainder were fitted as doublets. Each collected data set (sinogram), which
represents a temporal/thermal frame, consists of 33 rotational and 43 translational steps, while
each back-projected data set consists of 1849 (43$\times$43) patterns. An acquisition time of about
0.4 second per measurement (i.e. for particular translational position and rotational orientation)
gives an overall collection time of approximately 10 minutes to record a single frame.

\begin{figure} [!b]
\centering
\includegraphics
[scale=0.8] {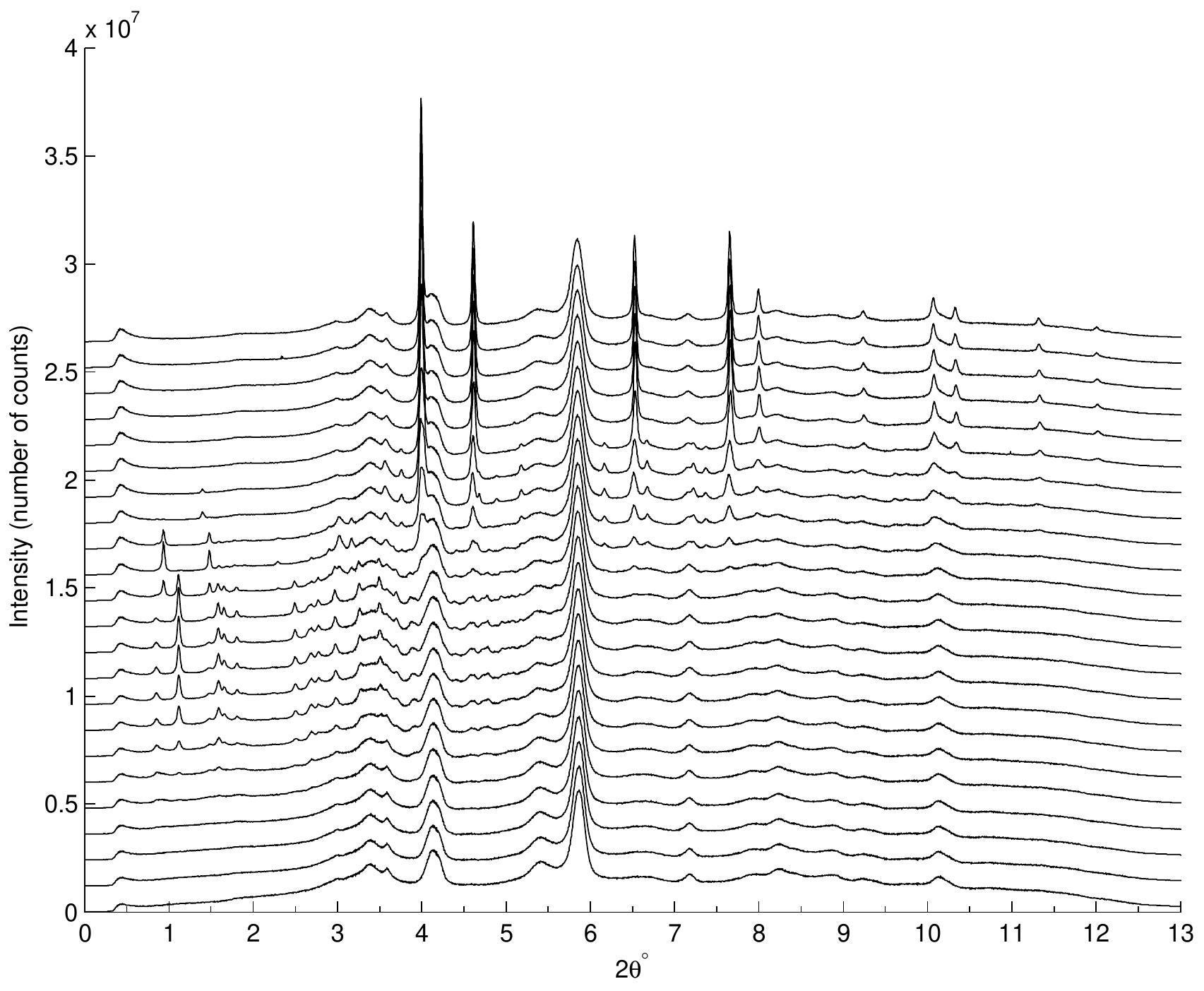}
\caption[Stack plot of the 23 data sets of nickel chloride.]%
{Sums of back-projected diffraction patterns of the 23 data sets of nickel chloride presented in a
stack plot. The curves in bottom-up order correspond to the data sets in their temperature and
temporal order.} \label{NiCl2ndStackBP1}
\end{figure}

\begin{figure} [!h]
\centering
\includegraphics
[scale=2.0] {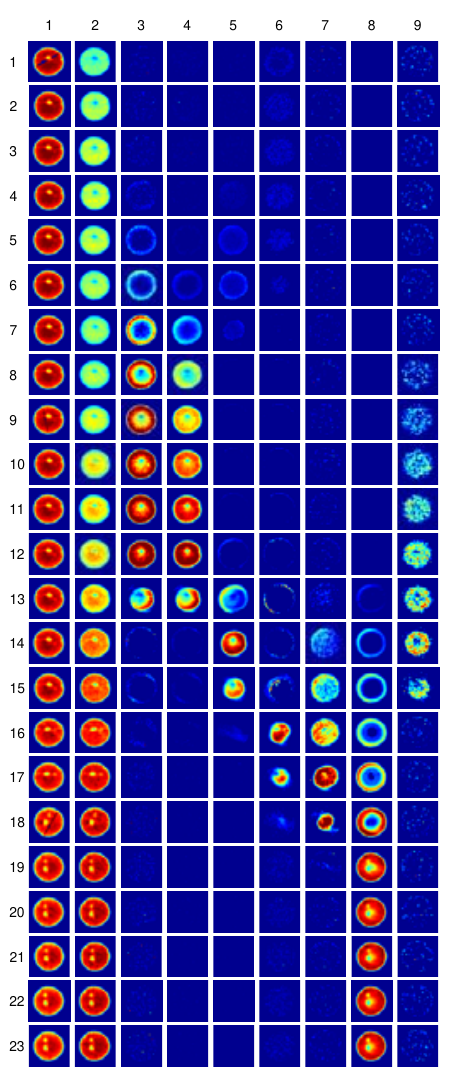}
\caption[Phase distribution patterns of nickel chloride data.]%
{Phase distribution patterns of nickel chloride data. The rows represent time/temperature frames
while the columns represent chemical/crystallographic phases. The columns (from left to right)
stand for $\gamma$-alumina, suspected transitional form of alumina, NiCl$_2$(en)(H$_2$O)$_4$
precursor, possible anhydrated form of the precursor, Ni$_2$Cl$_2$(en), NiCl$_2$-[R3-MH], HCP
Ni-[P63/MMC], FCC Ni-[FM3-M], and an unknown phase.} \label{NiCl2ndMask1}
\end{figure}

The stack plot of the sums of back-projected diffraction spectra is shown in Figure
\ref{NiCl2ndStackBP1}. A number of phase distribution patterns (PDP) have been observed with the
main ones being displayed in Figure \ref{NiCl2ndMask1}. The idea of the PDPs is to group the phases
in their spatial distribution in the frames as presented in the 2D color-coded tomographic images.
By classifying each particular sequence of tomographic images into a particular PDP and matching
the energies/wavelengths for which that PDP is obtained to a standard diffraction pattern, the
phase can be identified, and hence its spatial and temporal evolution is exposed. The stack plot of
the sums plays an assistant role in this process. The general assumption is that each PDP is a
finger print of a single crystallographic entity.

\begin{figure} [!h]
\centering
\includegraphics
[scale=1.2] {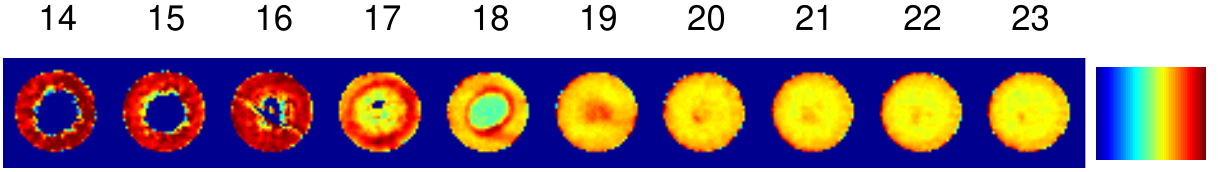}
\caption[Crystallite size distribution of FCC nickel.]%
{Crystallite size distribution of FCC nickel by mapping the FWHM of the (220) reflection at $2\SA =
6.51^\circ$. The color bar ranges between 0$^\circ$ (blue) and 0.065$^\circ$ (red). The blue maps
the regions where no FCC nickel peaks exist and hence the FCC phase is absent, while the red
represents the regions of smallest crystal size and the yellow represents the regions of largest
crystal size. The numbers shown on the top are the indices of the data sets.} \label{fwhm}
\end{figure}

Several chemical and crystallographic phases have been identified definitely or tentatively. These
include $\gamma$-alumina, a suspected transitional form of alumina, nickel chloride ethylenediamine
tetrahydrate [NiCl$_2$(en)(H$_2$O)$_4$] precursor, nickel chloride ethylenediamine
[Ni$_2$Cl$_2$(en)], nickel chloride (NiCl$_2$), hexagonal close packed (HCP) nickel, and face
centered cubic (FCC) nickel. Crystallite size growth for the FCC nickel has also been monitored and
mapped using a Scherrer-based analysis, as seen in Figure \ref{fwhm}. Standard diffraction patterns
obtained from the Inorganic Crystal Structure Database (ICSD) \cite{ICSD} have been used for phase
identification. The full details of this study can be found in \cite{SochiThesisBBK2010}.

It should be remarked that the whole operation of processing and analyzing nickel chloride data
(which involves converting tens of thousands of 2D binary images to 1D numeric diffraction
patterns; obtaining the sums of the stack plot; aligning, visualizing and back projecting the
sinograms; curve-fitting several millions of peaks; phase identification; and final visualization)
has been completed in just a few days. This stands in a sharp contrast to the required time and
effort, estimated to be weeks or even months of hard work, to perform such a task manually using
conventional tools.

\section{Conclusions} \label{Conclusions}

The EasyDD project is one step in the right direction for the future development of computational
tools to deal with the growing demand on data processing and analysis capabilities. Due to the
recent developments in the technology of radiation sources and data acquisition systems, this
approach is an important endeavor in developing software that can cope with the massive and
ever-increasing size of data collections that are generated in modern multi-tasking scientific
experiments. EasyDD has already proved to be a crucial component within a number of key studies.
Without the efficiency and speed of EasyDD, and without its effective strategies, such as the
multi-batch processing approach, some of these studies may not have happened.

During the development and use of EasyDD, a novel approach for processing and analyzing massive
spectral data collections has emerged in which phase distribution pattern diagrams, combined with
stack plots and standard diffraction patterns from powder diffraction databases, play a key role in
summarizing and presenting huge data sets in manageable form as an essential step to identifying
phases and tracking the evolution of complex physical and chemical systems. As such, the
developmental studies represented by the nickel chloride show case take diffraction-imaging
capabilities way beyond those of previous landmark studies, such as Bleuet \etal\
\cite{BleuetWDSHW2008} and Espinosa-Alonso \etal\ \cite{EspinosaOJBJe2009}, particularly for {\it
in situ} and dynamic phase transformation studies. It is anticipated that these developments should
have a significant future impact, at the scientific, technological and industrial level, within
several fields of research such as catalysis, {\it in operando} studies, phase transformations,
dynamic stress imaging of construction and biological materials, etc.

In the process of analyzing nickel chloride system, as outlined in the case study, several chemical
and crystallographic phases have been identified and mapped spatially and temporally, thereby
leading to important scientific implications. This system emerges as an outstanding example of what
can be achieved in following its evolution (precursor, intermediates and final active phase) in
terms of time, temperature, crystallite size and spatial distribution. The high level of detail
extracted enabled us to elucidate the overall evolving chemistry while also revealing new
information on the physical state of the catalyst and providing evidence and suggesting a mechanism
for the dynamic development. This particular study stands out as a vivid example that demonstrates
the capabilities and potential of these X-ray imaging and analysis techniques.

\clearpage
\section{Acknowledgement and Statement} \label{Acknowledgements}

The author would like to thank the Engineering and Physical Sciences Research Council (EPSRC) of
the United Kingdom for partial funding of this research. The author also acknowledges valuable
contributions from Prof. Paul Barnes, Dr. Simon Jacques and Dr. Olivier Lazzari to the EasyDD
project. The latest version of EasyDD with its documentation can be obtained free-of-charge from
the author.

\clearpage
\renewcommand{\refname}{}

\clearpage

\section{References} \label{bibliography}

\vspace{-2.1cm}

\begin{thebibliography}{3}

\bibitem[{(2007)}]{BealeJBBW2007}
Beale A.M., Jacques S.D.M., Bergwerff J.A., Barnes P. and Weckhuysen B.M. (2007) Tomographic Energy
Dispersive Diffraction Imaging as a Tool To Profile in Three Dimensions the Distribution and
Composition of Metal Oxide Species in Catalyst Bodies. Angewandte Chemie 119(46): 8988-8991.

\bibitem[{(2008)}]{BleuetWDSHW2008}
Bleuet P., Welcomme E., Dooryh\'{e}e E., Susini J., Hodeau J-L. and Walter P. (2008) Probing the
structure of heterogeneous diluted materials by diffraction tomography. Nature Materials 7:
468-472.

\bibitem[{(2010)}]{CernikHMPAe2010}
Cernik R.J., Hansson C.C.T, Martin C.M., Preuss M., Attallah M., Korsunsky A.M., Belnoue J.P., Jun
T.S., Barnes P., Jacques S., Sochi T. and Lazzari O. (2011) A synchrotron tomographic
energy-dispersive diffraction imaging study of the aerospace alloy Ti 6246. Journal of Applied
Crystallography 44(1): 150-157.

\bibitem[{(2009)}]{Diamond}
Diamond Light Source: \url{www.diamond.ac.uk}.

\bibitem[{(2009)}]{EspinosaOJBJe2009}
Espinosa-Alonso L., O'Brien M.G., Jacques S.D., Beale A.M., de Jong K.P., Barnes P. and Weckhuysen
B.M. (2009) Tomographic energy dispersive diffraction imaging to study the genesis of Ni
nanoparticles in 3D within $\gamma$-Al$_2$O$_3$ catalyst bodies. Journal of the American Chemical
Society 131(46): 16932-16938.

\bibitem[{(2009)}]{ESRF}
European Synchrotron Radiation Facility (ESRF): \url{www.esrf.eu/}.

\bibitem[{(2013)}]{GibsonZJBCe2013}
Gibson E., Zandbergen M.W., Jacques S., Biao C., Cernik R., O'Brien M.G., Di Michiel M., Weckhuysen
B.M. and Beale A.M. (2013)  Non-invasive Spatiotemporal Profiling of the Processes of Impregnation
and Drying within Mo/Al2O3 Catalyst Bodies by a Combination of X-ray Absorption Tomography and
Diagonal Offset Raman Spectroscopy. DOI: 10.1021/cs300746a.

\bibitem[{(2010)}]{ICSD}
Inorganic Crystal Structure Database: \url{http://icsd.ill.eu/icsd}.

\bibitem[{(2010)}]{HEXITEC}
Jones L., Seller P., Wilson M. and Hardie A. (2009) HEXITEC ASIC - a pixellated readout chip for
CZT detectors. Nuclear Instruments and Methods in Physics Research Section A 604(1-2): 34-37.

\bibitem[{(2011)}]{JacquesMBSOe2011}
Jacques S.D.M, Di Michiel M., Beale A.M., Sochi T., O'Brien M.G., Espinosa-Alonso L., Weckhuysen
B.M. and Barnes P. (2011) Dynamic X-Ray Diffraction Computed Tomography Reveals Real-Time Insight
into Catalyst Active Phase Evolution. Angewandte Chemie 50(43): 10148-10152.

\bibitem[{(2009)}]{LazzariJSB2009}
Lazzari O., Jacques S., Sochi T. and Barnes P. (2009) Reconstructive colour X-ray diffraction
imaging - a novel TEDDI imaging method. Analyst, 134(9): 1802-1807.

\bibitem[{(2012)}]{LazzariEJSMe2012}
Lazzari O., Egan C.K., Jacques S.D.M., Sochi T., Di Michiel M., Cernik R.J. and Barnes P. (2012) A
new approach to synchrotron energy-dispersive X-ray diffraction computed tomography. Journal of
Synchrotron Radiation, 19(4): 471-477.

\bibitem[{(2012)}]{ObrienJMBWB2012}
O'Brien M.G., Jacques S.D.M., Di Michiel M., Barnes P., Weckhuysen B.M. and Beale A.M. (2012)
Active phase evolution in single Ni/Al$_2$O$_3$ methanation catalyst bodies studied in real time
using combined $\mu$-XRD-CT and $\mu$-absorption-CT. Chemical Science, 3(2): 509-523.

\bibitem[{(2010)}]{SochiThesisBBK2010}
Sochi T. (2010) High Throughput Software for Powder Diffraction and its Application to
Heterogeneous Catalysis. PhD thesis, Birkbeck College London.

\bibitem[{(2011)}]{SochiEDFPaper2011}
Sochi T. (2011) Computational techniques for efficient conversion of image files from area
detectors. Sensors and Actuators A: Physical, 168(1): 72-76.

\bibitem[{(2011)}]{WilsonBCHJe2011}
Wilson M.D., Barnes P., Cernik R.C., Hansson C.C.T., Jacques S., Jones L.L., Seller P., Sellin
P.J., Sochi T., Veale M.C., Veeramani P., Withers P.J. and Youd C.P. (2010) Comparison of the X-Ray
Performance of Small Pixel CdTe and CZT Detectors. Nuclear Science Symposium Conference Record,
pages 3942-3946.

\end{thebibliography}
\end{document}